# Lattice-supported surface solitons in nonlocal nonlinear media

Yaroslav V. Kartashov, Victor A. Vysloukh*, and Lluis Torner

*ICFO-Institut de Ciencies Fotoniques, and Universitat Politecnica de Catalunya,*

*Mediterranean Technology Park, 08860 Castelldefels (Barcelona), Spain*

We reveal that lattice interfaces imprinted in nonlocal nonlinear media support surface solitons that do not exist in other similar settings, including interfaces of local and nonlocal uniform materials. We show the impact of nonlocality on the domains of existence and stability of the surface solitons, focusing on new types of dipole solitons residing partially inside the optical lattice. We find that such solitons feature strongly asymmetric shapes and that they are stable in large parts of their existence domain.

Nonlinear surface waves localized at the very interface between two optical media continuously attract attention because of their rich physical properties [1-3]. However, the high light intensity levels required for their excitation at interfaces of natural materials prevented their experimental observation in many cases. Recently, in a landmark advance, surface waves were observed experimentally in engineered interfaces made between a uniform medium and a waveguide array in AlGaAs [4,5]. Actually, such structures are a particular case of a general type of interfaces formed at the edge of a semi-infinite optical lattice imprinted in a nonlinear medium, which support novel types of surface solitons [6,7]. In this Letter we address the impact of a nonlocal nonlinearity on the properties of surface solitons supported by optical lattice interfaces.

We thus study nonlinear surface waves supported by a semi-infinite optical lattice imprinted in a nonlocal nonlinear medium. Nonlocality of the nonlinear response is a characteristic feature of many nonlinear materials including semiconductors, liquid crystals and photorefractive materials [8-10]. Nonlocality is known to drastically alter the soliton properties (see [11] for an overview), for example by having a strong stabilizing action [12]. Remarkably, nonlocality results in the formation of stable complex solitons, such us multi-pole solitons [13-17]. Here we reveal that a nonlocal nonlinearity makes possible the existence of a new type of surface solitons at lattice interfaces that do not exist at other physical settings, such us local interface or interface of nonlocal uniform materials. We find dipole surface lattice solitons that turn out to be strongly asymmetric but still stable, keeping their internal structure upon propagation even in the presence of considerable broadband perturbations.

We consider propagation of a paraxial laser beam along the $\xi$ axis at the interface of semi-infinite periodic lattice imprinted in a nonlocal nonlinear medium. Light propagation is described by the system of equations for the field amplitude $q$ and the nonlinear contribution to refractive index $n$:

$$i\frac{\partial q}{\partial \xi} = -\frac{1}{2}\frac{\partial^2 q}{\partial \eta^2} - qn - pRq,$$
$$n - d\frac{\partial^2 n}{\partial \eta^2} = |q|^2, \tag{1}$$

where $\eta, \xi$ stand for the normalized transverse and longitudinal coordinates, respectively; $p$ is the lattice depth; the function $R(\eta) = 0$ at $\eta < 0$ and $R(\eta) = 1 - \cos(\Omega\eta)$ at $\eta \geq 0$ describes the profile of a semi-infinite lattice with frequency $\Omega$; the parameter $d$ stands for the degree of nonlocality of the nonlinear response. Such lattices can be imprinted, e.g. in liquid crystals, by applying a proper voltage to the boundaries of elongated cells, as demonstrated in Ref. [9]. In such systems, nonlocality appears due to reorientational nonlinearity (i.e., a variation of the orientation of the liquid crystal molecules in one location causes a reorientation of the molecules in other parts of the crystal) and its degree can be tuned externally [8]. Notice that an applied voltage may also cause periodic variation in the nonlocality degree at $\eta > 0$: however, for the sake of generality here we consider the simplified model (1). Semiconductors possessing nonlocal nonlinearities are promising candidates for fabrication of lattice interfaces, too. The system (1) describes a local nonlinear response at $d \to 0$ and a strongly nonlocal response at $d \to \infty$. Similar material equations were used to study surface waves at the interface of uniform diffusive media and it was shown that nonlocality increases the energy threshold for surface wave formation and in general makes their excitation more difficult, especially when diffusion occurs through the interface [18,19].

To find the amplitude profiles of lattice-supported surface solitons we let $q(\eta,\xi) = w(\eta)\exp(ib\xi)$, where $w(\eta)$ is a real function and $b$ is the real propagation constant. The profiles of the soliton solutions with $w, n \to 0$ at $\eta \to \pm\infty$ were obtained numerically. To test their stability we searched for perturbed solutions in the form $q = (w + u + iv)\exp(ib\xi)$, where $u(\eta,\xi)$ and $v(\eta,\xi)$ are the real and the imaginary parts of perturbations that can grow with a complex rate $\delta$ upon propagation. The perturbation profiles and growth rates $\delta$ are obtained from the linear eigenvalue problem

$$\begin{aligned}
\delta u &= -\frac{1}{2}\frac{d^2 v}{d\eta^2} + bv - nv - pRv, \\
\delta v &= \frac{1}{2}\frac{d^2 u}{d\eta^2} - bu + nu + w\Delta n + pRu,
\end{aligned} \qquad (2)$$

where $\Delta n = 2\int_{-\infty}^{\infty} G(\eta-\lambda)w(\lambda)u(\lambda)d\lambda$ and $G(\eta) = (1/2d^{1/2})\exp(-|\eta|/d^{1/2})$ stands for the response function of the nonlocal medium.

We found that the simplest ground-state surface solitons supported by the lattice interface reside in the first lattice channel, where the field amplitude reaches its maximum value (see Figs. 1(a) and 1(b)). To illustrate the properties of the solitons, we set $\Omega = 2$, and vary $b$, $p$, and $d$. The energy flow $U = \int_{-\infty}^{\infty}|q|^2 d\eta$ of the simplest solitons monotonically increases with $b$ everywhere except for a narrow region close to the cutoff $b_{\text{low}}$ for soliton existence, where the derivative $dU/db$ becomes negative (this region is not even visible in the $U(b)$ dependence depicted in Fig. 2(a)). As $b \to b_{\text{low}}$ solitons penetrate deeper inside the lattice and oscillations on their profile become more pronounced. For large values of $b$, the soliton energy concentrates almost completely in the first lattice channel. In the regime of strong nonlocality the width of the refractive index profile substantially exceeds the soliton width, and hence there are almost no oscillations on the $n(\eta)$ curve. The cutoff $b_{\text{low}}$ monotonically increases with $p$ (see Fig. 2(b)), and only slightly increases with the nonlocality degree $d$. As in the case of interface between uniform diffusive media [18], increasing the nonlocality strength results in an increase of the threshold surface wave energy $U_{\min}$. Thus, at $p=1$ one has $U_{\min} = 0.409$ for $d = 0.5$ while one gets $U_{\min} = 1.166$ for $d = 5$. Ground-state surface solitons are stable in the entire existence domain where $dU/db > 0$; they are unstable only in a narrow region near the existence cutoff where $dU/db \leq 0$.

The central finding reported in this Letter is that lattice interfaces imprinted in nonlocal media support *dipole solitons,* comprising two out-of-phase poles that reside at *different* sides of the interface (see Fig. 1(c) and 1(d)). The poles of such solitons are glued together because in nonlocal media the correction to the refractive index is determined by the light intensity distribution in the entire transverse plane, so that equilibrium configurations of out-of-phase beams are possible for a proper separation between them. Due to the presence of the interface dipole surface solitons are strongly asymmetric: multiple oscillations develop on the soliton profile at $\eta > 0$, while at $\eta < 0$ the field distribution is smooth. The asymmetry of the dipole solitons increases with lattice depth and with the nonlocality degree. Importantly, note that such solitons *do not exist* at interfaces of an optical lattice and a uniform medium with local nonlinearity. We did not find such solutions in interfaces made in uniform media, i.e., latticeless nonlocal media where the refractive index profile is described by a step-like function $R(\eta)$. Namely, our code did not converge to a solution residing at the interface for any of the numerous trial functions that we used. Thus, we conclude that the lattice is a necessary ingredient for the existence of the dipole surface solitons described here.

We found that dipole surface lattice solitons exhibit lower $b_{\text{low}}$ and upper $b_{\text{upp}}$ cutoffs for existence. As $b \to b_{\text{low}}$ the asymmetry of the soliton poles increases. This is because the right soliton pole substantially expands into the lattice region, while the left pole moves deeper into the uniform medium (Fig. 1(c)). Still, even small-amplitude right poles keep high-amplitude left poles glued to the interface. At $b \to b_{\text{upp}}$ dipole surface solitons become less asymmetric (Fig. 1(d)). The energy flow is a monotonically increasing function of $b$ except for narrow regions near both lower and upper cutoffs where $dU/db \leq 0$ (Fig. 2(c)). The domain of existence of dipole solitons gradually shrinks with increase of lattice depth (Fig. 2(d)). In contrast, the upper cutoff first decreases and then increases with $d$, resulting in a substantial expansion of the existence domain in strongly nonlocal media (Fig. 2(e)). This is because the lower cutoff (not shown) changes with $d$ only slightly (thus, at $p = 1$ one has $b_{\text{low}} \approx 1.3$).

A detailed linear stability analysis of the above stationary solutions revealed that dipole surface solitons can be stable in large part of their existence domain. Stabilization occurs below a critical propagation constant value $b_{cr}$. Dipoles are stable for $b_{low} < b < b_{cr}$ and oscillatory unstable for $b_{cr} \leq b < b_{upp}$. Figure 2(f) shows the dependence of the perturbation growth rate on $b$. With an increase of the lattice depth or the nonlocality degree, $b_{cr}$ gradually approaches $b_{upp}$ (see Fig. 2(d) and 2(e)). Notice that in the strongly nonlocal regime a narrow instability domain emerges also near the lower cutoff. The new domain gradually expands with increasing $d$. Direct numerical simulations of Eq. (1) with input conditions $q|_{\xi=0} = w(1 + \rho)$, where $\rho(\eta)$ is a broadband random perturbation, always confirmed the results by the linear stability analysis. Illustrative propagations of perturbed stable fundamental and dipole surface solitons are shown in Fig. 3. Notice that strong perturbation can cause small oscillations of the distance between the poles of the dipole solitons, as observable in Fig. 3(b). However, the surface lattice solitons remain robust.

We thus summarize by stressing that we reported on the existence of new type of complex surface solitons supported by the interface of a periodic, finite optical lattice imprinted in a nonlocal nonlinear medium. Our findings indicate that nonlocality and the refractive index modulation in the lattice region are necessary ingredients for the existence of surface dipoles that remain stable in spite of their strong shape asymmetry.

*Also with Universidad de las Americas, Puebla, Mexico.

# References with titles

# References without titles

# Figure captions

Figure 1 (color online). Profiles of single surface solitons with $b = 1.34$ (a) and $b = 2$ (b) at $p = 1$, $d = 5$. Profiles of dipole surface solitons with $b = 1.3$ (c) and $b = 3.2$ (d) at $p = 1$, $d = 3$. Black curves show field distribution, red curves show refractive index profile. In gray regions $R(\eta) \geq 1$, while in white regions $R(\eta) < 1$.

Figure 2. (a) Energy flow vs propagation constant for the single surface solitons at $p = 1$. (b) Domain of existence for single surface solitons on $(b, p)$ plane at $d = 1$. (c) Energy flow vs propagation constant for dipole surface solitons at $p = 1$. Domains of existence and stability for dipole surface solitons on $(b, p)$ plane at $d = 1$ (d) and on $(b, d)$ plane at $p = 1$ (e). (e) Real part of perturbation growth rate vs propagation constant for dipole surface soliton at $p = 1.25$ and $d = 1$. Points marked by circles in (a) and (c) correspond to the soliton profiles shown in Figs. 1(a),1(b) and 1(c),1(d), correspondingly.

Figure 3. Propagation dynamics of single surface soliton at $b = 1.4$, $p = 1$, $d = 5$ (a) and dipole surface soliton at $b = 1.6$, $p = 1$, $d = 3$ (b). The field modulus is shown. In both cases white noise with variance $\sigma_{noise}^2 = 0.01$ was added to the input field distributions.

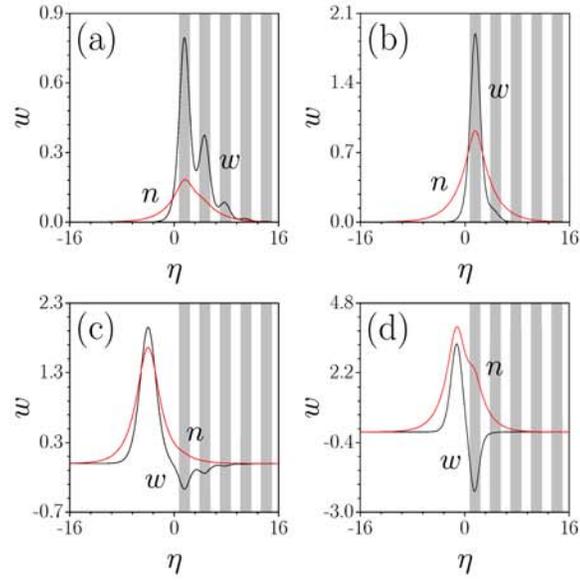

Figure 1 (color online). Profiles of single surface solitons with $b=1.34$ (a) and $b=2$ (b) at $p=1$, $d=5$. Profiles of dipole surface solitons with $b=1.3$ (c) and $b=3.2$ (d) at $p=1$, $d=3$. Black curves show field distribution, red curves show refractive index profile. In gray regions $R(\eta) \geq 1$, while in white regions $R(\eta) < 1$.

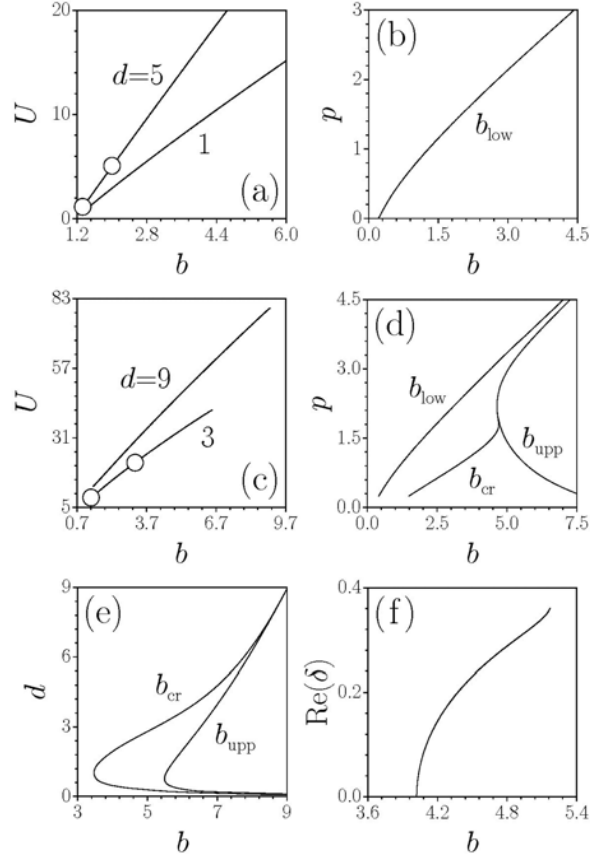

Figure 2. (a) Energy flow vs propagation constant for the single surface solitons at $p=1$. (b) Domain of existence for single surface solitons on $(b,p)$ plane at $d=1$. (c) Energy flow vs propagation constant for dipole surface solitons at $p=1$. Domains of existence and stability for dipole surface solitons on $(b,p)$ plane at $d=1$ (d) and on $(b,d)$ plane at $p=1$ (e). (e) Real part of perturbation growth rate vs propagation constant for dipole surface soliton at $p=1.25$ and $d=1$. Points marked by circles in (a) and (c) correspond to the soliton profiles shown in Figs. 1(a),1(b) and 1(c),1(d), correspondingly.

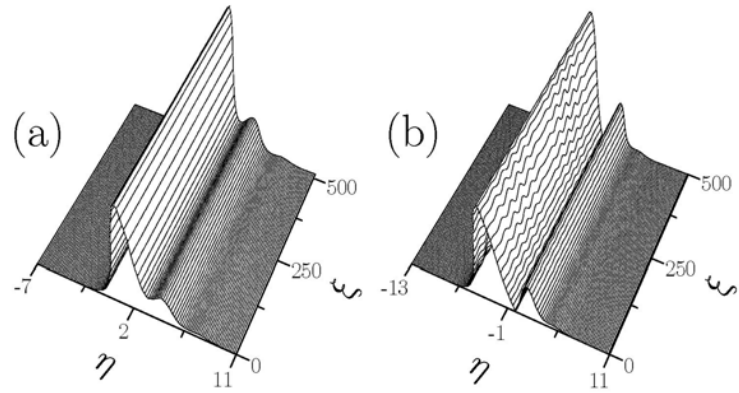

Figure 3. Propagation dynamics of single surface soliton at $b = 1.4$, $p = 1$, $d = 5$ (a) and dipole surface soliton at $b = 1.6$, $p = 1$, $d = 3$ (b). The field modulus is shown. In both cases white noise with variance $\sigma^2_{\text{noise}} = 0.01$ was added to the input field distributions.